\documentclass[aps,prstb,preprint,onecolumn,showpacs,groupedaddress,]{revtex4-1}
\usepackage{graphicx}
\usepackage{epstopdf}
\usepackage{dcolumn}
\usepackage{bm}
\usepackage{amssymb}
\usepackage{amsmath}
\begin{document}

\title{Analysis of $\boldmath{\check{\text{C}}}$erenkov free-electron lasers}
\author{Yashvir Kalkal}
\email{yashvirkalkal@gmail.com}
\author{Vinit Kumar}
\email{vinit@rrcat.gov.in}

\affiliation{Accelerator and Beam Physics Laboratory, Raja Ramanna Centre for Advanced Technology, Indore, MP 452013, India}

\begin{abstract}
We present an analysis of a $\check{\text{C}}$erenkov free-electron laser (FEL) driven by a flat electron beam. In this system, an electron beam travelling close to a dielectric slab placed at the top of an ideal conductor interacts with the co-propagating  electromagnetic surface mode. The surface mode arises due to singularity in the reflectivity of the dielectric slab for the incident evanescent wave. Under suitable conditions, the surface mode grows as a result of interaction with the electron beam. We show that the  interaction of the surface mode with the co-propagating electron beam can be rigorously understood by analyzing the singularity in the reflectivity. Using this approach, we set up coupled Maxwell-Lorentz equations for the system, in analogy with conventional undulator based FELs. We solve these equations analytically in the small signal regime to obtain formulae for the small signal gain, and the spatial growth rate. Saturation behaviour of the system is analyzed by solving these equations numerically in the nonlinear regime. Results of numerical simulations are in good agreement with the analytical calculations in the linear regime. We find that $\check{\text{C}}$erenkov FEL under appropriate conditions can produce copious coherent terahertz (THz) radiation.
\end{abstract}

\maketitle

\section{\label{sec:level1}Introduction}

An electron moving in a close proximity to a dielectric material emits $\check{\text{C}}$erenkov radiation \cite{Cerenkov} with angle of emission given by 
\begin{equation}
\cos\theta=\frac{1}{\beta \sqrt{\epsilon \mu}}~.
\end{equation}
Here, $\epsilon$ and $\mu$ are relative permittivity and relative permeability respectively of the dielectric medium; $\beta = v/c$, $v$ is the electron's speed and $c$ is the speed of light in vacuum. In 1947, Ginzburg \cite{Ginzburg} proposed that this effect can be utilised to make a source of electromagnetic radiation. Since then many experimental and theoretical investigations have been made to understand the generation of radiation from such sources \cite{Walsh2,Walsh3,Garate2,Tripathi1,Bhasin,Fares1,Fares2,Kheiri,Ganguly,Freund,fuente1,Fuentethesis,Walsh1,Owens1,Brau1,Li1, Sharma,Li3, Asgek1,Asgekar2,Asgekar3,Danos1,Seo,Owens2,Li2,Garate1}. During recent times, there is a demand for powerful, compact and tunable sources of electromagnetic radiation in the terahertz (THz) regime for numerous applications in research and industry \cite{Seigel,Clery,THz}. $\check{\text{C}}$erenkov free-electron laser (CFEL), which uses a low-energy electron beam ($\sim$30 keV) will be a promising source of the THz radiaton due to its compact size \cite{Owens2}.

In a CFEL, coherent electromagnetic radiation is produced due to the interaction of an electron beam with the co-propagating surface mode in a dielectric based slow wave structure. The slow wave structures investigated so far are: single or double dielectric slab in the planar geometry, and cylindrical waveguide lined with a dielectric material. Walsh {\textit{et al.}} \cite{Walsh2,Walsh3} made one of the earlier theoretical analysis for the small-signal gain in CFEL and compared with the gain of conventional undulator based FEL. They used linearized Vlasov equation to evaluate the modulation in the current density due to interaction with the electromagnetic field. The modulated current density was then used to calculate the energy gain of the CFEL. In their analysis, they neglected the space charge effect and performed the calculations only in the low-gain regime. The analysis was further extended by several authors \cite{Garate2,Tripathi1,Bhasin,Kheiri,Fares1,Fares2} by setting up the coupled Maxwell-Vlasov equations, and the growth rate was calculated. In most of the earlier analyses, the electron motion was assumed to be one dimensional in the presence of strong external static magnetic field \cite{Walsh1,Walsh2,Walsh3,Garate2}. Freund and Ganguly \cite{Freund,Ganguly} developed a three-dimensional (3D) theory for the CFEL in the cylindrical configuration, by setting up coupled Maxwell-Lorentz equations to evolve the electromagnetic field and the electron trajectories. Fuente {\textit{et al.}} \cite{fuente1,Fuentethesis} extended this theory by including the loss due to liner fluctuations and reported the successful operation of such devices.  

Another approach, known as the hydrodynamic approach \cite{Walsh1,Li3,Owens1,Brau1,Li1,Sharma}, has also been used by several authors to evaluate the dispersion relation and growth rate in the $\check{\text{C}}$erenkov FELs. In this approach, one treats the electron beam as a plasma dielectric and solves the Maxwell equations to find the dispersion relation of the system. The dispersion relation can be expanded in the Taylor series about the roots of no-beam dispersion to find the growth rate of electromagnetic field. Using this approach, Owens and Brownell \cite{Owens1} performed two-dimensional (2D) analysis for a CFEL based on single slab geometry. Assuming an infinite electron beam, they found the usual cubic dispersion relation for the system. Andrew and Brau \cite{Brau1} developed a 3D theory for the device and found that the gain reduces by an order of magnitude as compared to the 2D theory. The usual cubic dispersion relation gets replaced by the 5/2-power dispersion relation on accounts of 3D effects. The hydrodynamic approach was extended to the double-slab configuration by Li {\textit{et al.}} \cite{Li1} and corresponding asymmetric case has been studied by Sharma and Mishra \cite{Sharma}. This approach works well in the linear regime, but is difficult to extend in the non-linear regime.

Recently, in Refs. \cite{Asgek1,Asgekar2,Asgekar3}, the authors have discussed an approach based on Maxwell-Lorentz equations for the analysis of $\check{\text{C}}$erenkov FELs, which is similar to that of conventional FELs. Asgekar and Dattoli \cite{Asgekar2,Asgekar3} have used this approach to calculate the gain and saturation intensity in the $\check{\text{C}}$erenkov FELs, for the single slab geometry. In their analysis, they have however not included the variation of the electromagnetic fields in a direction perpendicular to the dielectric surface in a rigorous way, while solving the coupled Maxwell-Lorentz equations. In particular, they have not calculated the power in the electromagnetic field taking the transverse variation of the field into account. As a result, the formulae derived in these references do not have dependence on the height of the electron beam from the dielectric surface.

In all the above mentioned analyses, size of the electron beam is taken to be either very large or infinite. Since the mode supported by the dielectric surface is evanescent in the direction perpendicular to the surface and confined in a region very close to the dielectric surface, it is more appropriate to take a flat electron beam travelling very close to the dielectric surface to ensure significant interaction with the evanescent mode. We would like to emphasize that a flat beam has vertical size much smaller than its horizontal size over the entire interaction length. The first experimental observation of the $\check{\text{C}}$erenkov radiation by using flat electron beam was performed by Danos {\textit{et al.}} in 1953 \cite{Danos1}. In comparison to a round electron beam, flat beam with the same current allows more effective interaction with the surface mode, since all the electrons are at a reduced height from the dielectric surface. Also, a flat beam can allow much more current within the required dimension in the horizontal direction, and thus will help to enhance the output power of the device \cite{Seo}. Additionally, one can tune the operating frequency of device by varying the gap between the flat electron beam and the dielectric surface \cite{Owens1}.

In this paper, we analyse the $\check{\text{C}}$erenkov FEL driven by an infinitesimal thin flat electron beam in the single slab geometry, as shown in Fig.~1. We have set up the coupled Maxwell-Lorentz equations by appropriately taking into account the variation of the electromagnetic field in the vertical direction. Our approach incorporates the space charge effects, and it is easily extendible to the non-linear regime, unlike the approach based on coupled Maxwell-Vlasov equations. In this approach, the electromagnetic field due to a flat beam is presented as a spectrum of plane waves of different frequencies and having phase velocity equal to the electron beam velocity \cite{Toraldo}. These waves are evanescent in nature and decay, away from the electron beam. When the electron beam is kept sufficiently close to the dielectric medium, these evanescent waves are incident on the surface and give rise to reflected evanescent waves. Sum of the incident and reflected electromagnetic field effectively interacts with the co-propagating electron beam and gives rise to the coherent electromagnetic radiation, under suitable condition. The evaluation of reflectivity for a dielectric slab supported by an ideal conductor is easily done by satisfying the appropriate boundary conditions. For a given phase velocity of the evanescent wave, the reflection coefficient has a singularity at a particular frequency, which means that the structure supports a surface mode at that frequency \cite{Levi}. The interaction of the electron beam with this surface mode becomes the mechanism for the working of $\check{\text{C}}$erenkov FEL. By evaluating the dispersion relation for normal materials, we can see that the group velocity of surface mode in a CFEL is always positive. However, in case of negative-index material, the energy flows in backward direction \cite{Li3}. For this case, under certain conditions, the system develops a self feedback mechanism and can act like a backward wave oscillator (BWO). The device based on positive index materials will always have a travelling wave amplifier (TWA)-type interaction \cite{Brau1}. If we apply an external feedback mechanism to a low-gain CFEL, the system starts working as an oscillator. In this paper, we have studied this configuration of the $\check{\text{C}}$erenkov FEL in detail. We would like to point out that our approach here is similar to the one used by Kumar and Kim to analyze the working of Smith-Purcell free-electron laser (SP-FEL) \cite{VinitPRE,KimPRSTB,VinitPRSTB,VinitFEL05}. This approach was very successful for the detailed analysis of the system, and had removed inconsistency amongst different analyses of SP-FELs. Here, we are extending this analysis for the $\check{\text{C}}$erenkov FEL. However, unlike the case of SP-FEL described in their paper, the group velocity of the surface wave is positive.

In the next section, we set up the basic electromagnetic field equations for a single-slab geometry based CFEL driven by a flat electron beam. This is followed by the detailed calculation of singularity in the reflection coefficient of the dielectric surface. Next, we discuss the interaction of the surface mode with the electron beam and calculate the small-signal gain in Section III. In Sec. IV, we introduce collective variables to calculate the growth rate in small-signal high-gain regime. We extend our analysis to the non-linear regime by performing numerical simulations in Section V. Finally, in Sec. VI, we discuss the results and conclude our analysis.

\section{\label{sec:level2}BASIC ELECTROMAGNETIC FIELD EQUATIONS AND REFLECTIVITY ANALYSIS}
We start the analysis by setting up the electromagnetic fields due to a flat electron beam. We have closely followed the approach described in Ref. \cite{VinitPRE}, and extended it to the case of CFEL. The schematic of the system for a $\check{\text{C}}$erenkov FEL is shown in Fig. 1. The dielectric slab of thickness $d$, length $L$ and dielectric constant $\epsilon$ is supported on an ideal conductor. The flat electron beam is confined to move along the $z$ direction and at a height $h$ above the dielectric surface. We denote the electron speed by $v$. We are assuming here that the system has translational invariance in the $y$ direction.
\begin{figure}[h]
\includegraphics[width=14.5cm]{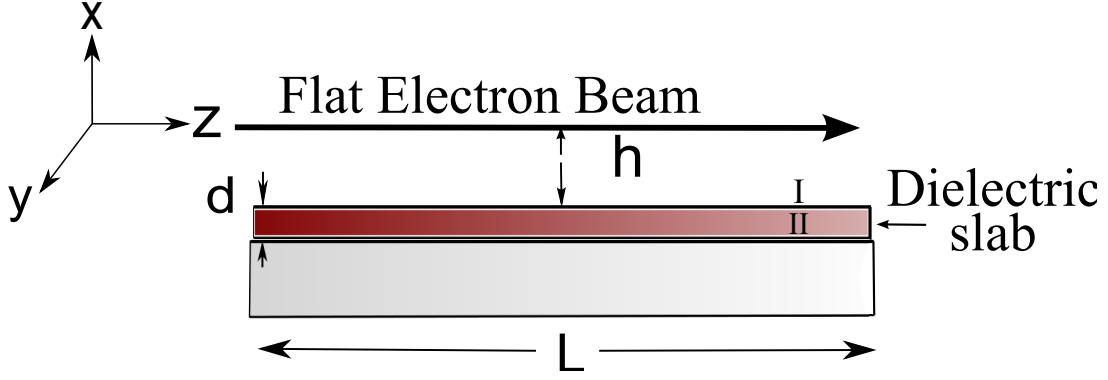}
\caption{Schematic of $\check{\text{C}}$erenkov FEL using a flat electron beam.}
\end{figure}

The current density for a flat beam has the form $K\delta(x)$. Here, $\delta(x)$ is the Dirac delta function and $K$ is the surface current density of electron beam, located at $x$=0. As discussed later in the section, the dielectric slab supports an evanescent surface mode at the resonant frequency of the system. Due to the interaction with the surface mode, the electron beam develops a modulation in the current density. The surface current density can be expanded in the Fourier series and the term at the resonant frequency $\omega$ will show the strongest interaction with the surface mode. We can then write the surface current density as  $K(z,t)e^{i(k_0z-\omega t)}+c.c.$, where $c.c.$ represents complex conjugate and $K(z,t)$=$(I/\Delta y)\langle e^{-i\psi} \rangle $. The beam current is represented as $I$, $\Delta y$ is the electron beam width in the $y$ direction, $\psi =k_0z-\omega t$ is the electron phase, $k_0$=$\omega/v$ and $\langle \cdots \rangle$ indicates averaging over the number of particles distributed over one wavelength of the evanescent mode. We assume that the surface current density will be a slowly varying function of the type $e^{\mu z}$ in $z$. Note that if the real part of $\mu$ is positive, the electrons become bunched at the resonant frequency. With this assumption, the surface current density can be finally written as $K_0e^{i(\alpha_0 z-\omega t)}+c.c.,$ where $\alpha_0$=$k_0-i\mu$ and $K_0$ is independent of $z$ and $t$.

The electromagnetic fields due to the above current density can be solved by using the Maxwell equations with appropriate boundary conditions. This gives us the following expression for the incident electromagnetic field \cite{VinitFEL05}:
\begin{eqnarray}
H_y^{\mathcal{I}}(x,z)=\frac{1}{2}\theta(x)K(z)\exp[-\theta(x)\Gamma x],
\end{eqnarray}  
where $\Gamma$=$(\alpha_0^2-\omega^2/c^2)^{1/2}$, $\theta(x)$=1 for $x>0$ and $\theta(x)$=-1 for $x<0$. The electromagnetic field is decaying away from the electron beam and has $e^{-i\omega t}$ time dependency. The flat electron beam acts as a source of the electromagnetic field. Now, due to the dielectric slab, this incident filed is reflected back towards the electron beam. The reflected and incident electromagnetic fields are coupled through the reflectivity $R$ of the dielectric surface. The reflected electromagnetic field is given as \cite{VinitFEL05}:
\begin{eqnarray}
H_y^{\mathcal{R}}(x,z)=-\frac{1}{2}K(z)R\exp[-\Gamma(2h+x)].
\end{eqnarray}  
The electromagnetic field has $H$ polarisation, which means that $H_x$=$H_z$=$E_y$=0. Using the Maxwell equation, $E_z$ can be written as $E_z$=$(i/{\epsilon_0\omega})(\partial H_y/\partial x-\delta(x)K)$. Now, the sum of incident and reflected electromagnetic field effectively interacts with the electron beam. The amplitude of the electromagnetic field experienced by the electron beam is obtained as \cite{VinitFEL05}:
\begin{eqnarray}
E_z(x=0,z)=\dfrac{iIZ_0}{2\beta\gamma \Delta y}(Re^{-2\Gamma h}-1)\langle e^{-i\psi}\rangle.
\end{eqnarray}
Here, $Z_0$=$1/(\epsilon_0 c)$= 377 $\Omega $ is the characteristic impedance of free space, $\gamma $ is the relativistic Lorentz factor and $\epsilon_0$ is the permittivity of free space. The total longitudinal electric field at the location of the electron beam has the form $E_ze^{i(k_0z-\omega t)}+c.c.$. 
\begin{table}[b]
\caption{\label{tab:table1}Parameters of a CFEL used in the calculation}
\begin{ruledtabular}
\begin{tabular}{lcdr}
~~~~~~Electron energy & 30 keV\\
~~~~~~Electron-beam height ($h$) & 35 $\mu$m\\
~~~~~~Electron-beam current ($I$) & 1 mA\\
~~~~~~Dielectric constant ($\epsilon$) &  13.1\\
~~~~~~Length of slab ($L$) & 0.15 m\\
~~~~~~Dielectric thickness ($d$) & 350 $\mu$m\\
~~~~~~Operating frequency & 0.1 THz\\
\end{tabular}
\end{ruledtabular}
\end{table}

\begin{figure}
\includegraphics[width=17.2cm]{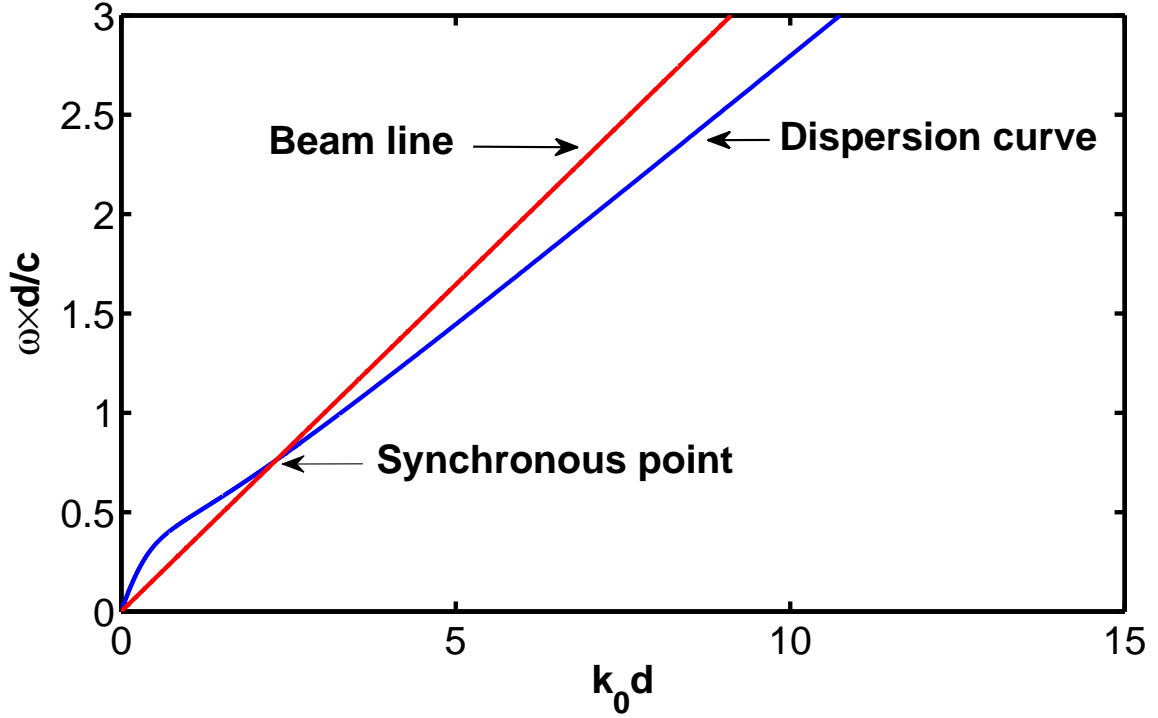}
\caption{Plot of the dispersion curve of the surface mode, and the Doppler line for the electron beam. At the intersection, we find the resonant frequency of the system. Parameters used in the calculation are given in Table I.}
\end{figure}
\begin{figure}
\includegraphics[width=15.0cm]{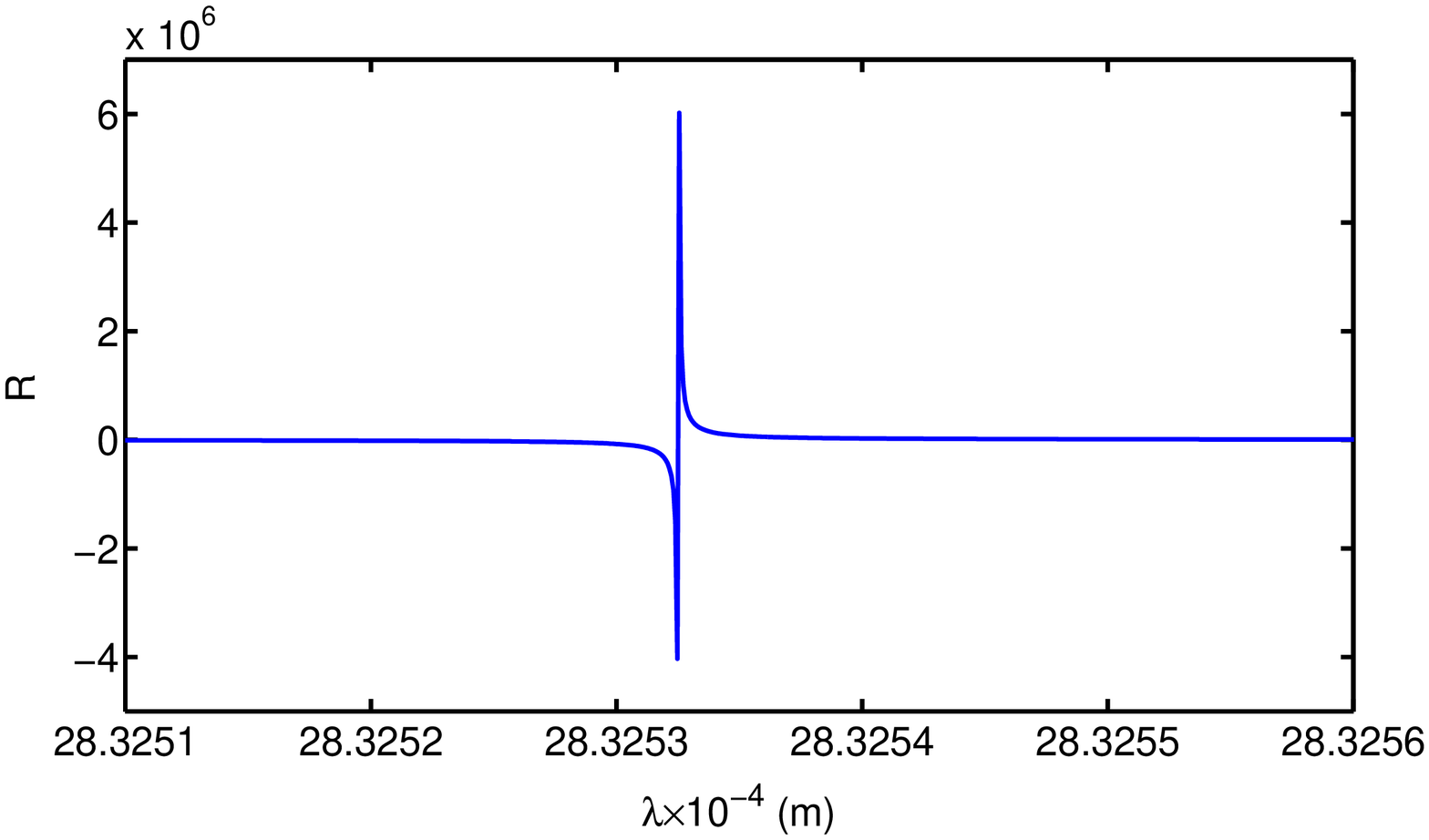}
\caption{Plot of $R$ as a function of wavelength, as calculated using Eq.~(5). Dielectric parameters used are taken from Table I, and $\beta$ = 0.33, corresponding to 30 keV electron beam. The singularity in $R$ appears at the resonant frequency of 0.1 THz}
\end{figure}

To calculate electromagnetic field in Eq. (4), we need to evaluate the reflectivity of the dielectric slab, supported by a metallic surface. By solving the Maxwell equations for the electromagnetic fields with appropriate boundary conditions, we find reflectivity of the dielectric slab as:
\begin{eqnarray}
R=\frac{1+r\tan (d\sqrt{\epsilon\beta^2k_0^2-\alpha_0^2})}{1-r\tan (d\sqrt{\epsilon\beta^2k_0^2-\alpha_0^2})},
\end{eqnarray}
where $r$=$\sqrt{\epsilon\beta^2k_0^2-\alpha_0^2}/\epsilon\sqrt{\alpha_0^2-\beta^2k_0^2}$. Note that for $\mu$=0, reflectivity has a singularity at $k_0$=$(1/b)\tan^{-1}(1/a)$, which is same as dispersion relation of the system as described in Refs. \cite{Walsh2,Asgek1,Brau1}. Here $a$=$(\gamma/\epsilon)\sqrt{\epsilon\beta^2-1}$ and $b$=$d\sqrt{\epsilon\beta^2-1}$. It has already been observed that the condition for a system to support surface mode of frequency $\omega$=$\beta ck_0$ is equivalent to requirement that the reflection coefficient is singular at that particular frequency \cite{Levi}. Figure 2 shows the plot of dispersion curve, and the reflectivity as a function of wavelength is plotted in Fig 3. The parameters used in our calculations are taken from the Dartmouth experiment \cite{Owens2,Brau1}, and are listed in Table~\ref{tab:table1}. For $\beta$=0.33, the singularity in $R$ occurs at 0.1 THz, which is the resonant frequency of the system, as shown in Fig. 2. A more careful observation of Eq. (5) indicates that the reflectivity is a function of frequency as well as the growth rate parameter $\mu$. This has important implication, while calculating the growth rate of the surface mode.

\begin{figure}[b]
\includegraphics[width=16.0cm]{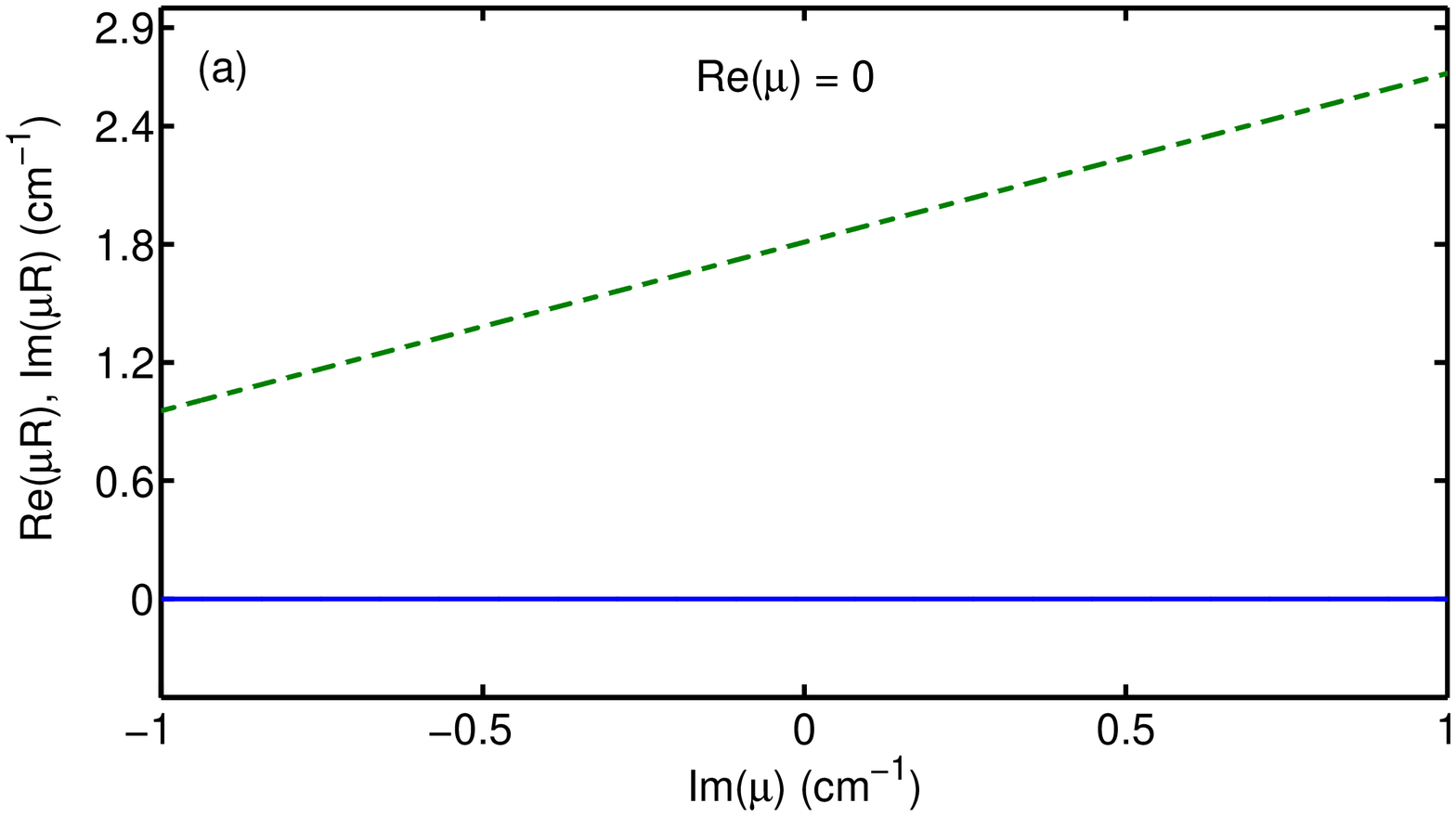}
\includegraphics[width=16.0cm]{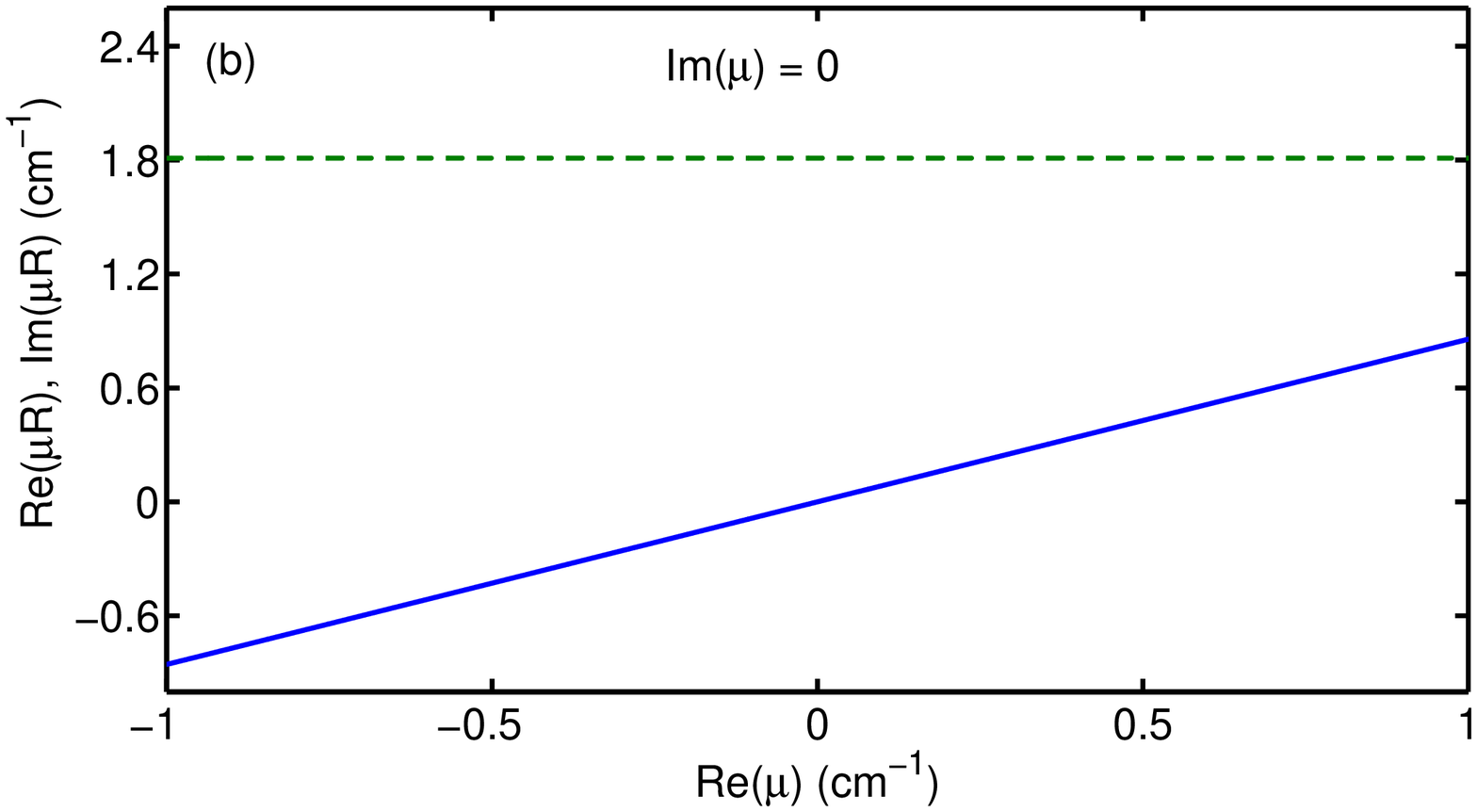}
\caption{Plots of imaginary (dashed) and real (solid) parts of $\mu R$ as a function of the imaginary (a) and real (b) parts of the growth rate parameter $\mu$ near the resonance frequency, i.e., 0.1 THz. By parametrising $R$ as $\frac{i\chi}{\mu}+\chi_1$ in this graph, we obtain $\chi$=1.81 per cm and $\chi_1$=0.86.}
\end{figure}

In order to study the nature of singularity in $R$ at $\mu$=0, we perform Laurent series expansion of $R$, as a function of $\mu$. By doing so, we obtain the following expression of $R$:
\begin{eqnarray}
R=\dfrac{m_0+m_1\mu+m_2\mu^2+o(\mu^3)....}{n_0+n_1\mu+n_2\mu^2+o(\mu^3)....},
\end{eqnarray}
where the coefficients of expansion are given as:
\begin{eqnarray}
n_1=\frac{-i}{k_0a b^2}\lbrack ad^2+ab^2\gamma^2+k_0 b d^2(1+a^2)\rbrack, ~~~~~~~~~
\end{eqnarray}
\begin{eqnarray}
n_2=\frac{-1}{2k_0^2b^4}\lbrack d^4+b^2d^2(1-2\gamma^2)+\gamma^2b^4(1-3\gamma^2)\rbrack ~~~~~~~~~~~~~~ \nonumber\\
~~~~+\frac{(1+a^2)}{2k_0a^2b^3}\lbrack ad^4+2k_0bd^4-ab^2d^2(1-2\gamma^2)\rbrack,~~~~~~~~~~~~~~
\end{eqnarray} 
$n_0$=0, $m_0$=2, $m_1$=$-n_1$ and $m_2$=$-n_2$. In Eq. (6), we have division of two infinite series. By performing the required algebra and keeping the terms of the order of $1/\mu$ and $\mu^0$, we obtain the following simple expression for the reflectivity:
\begin{eqnarray}
R=\frac{i\chi}{\mu}+\chi_1.
\end{eqnarray}
Here, $\chi$ and $\chi_1$ are given by:
\begin{eqnarray}
\chi=-im_0/n_1, \\
\chi_1=(m_1n_1-m_0n_2)/n_1^2~.
\end{eqnarray}
The parameters $\chi$ and $\chi_1$ are of paramount importance for any geometry of the $\check{\text{C}}$erenkov FEL. As described in the next section, parameter $\chi$ is associated with the growth rate of surface electromagnetic mode and $\chi_1$ is related to the space charge effect. For parameters listed in Table I, we find $\chi$=1.81 per cm and $\chi_1$=0.86 from Eqs. (10)-(11). We also confirmed these values by numerically evaluating the value of $\mu R$ and plotting it in the vicinity of $\mu$=0. The value of $\chi$ and $\chi_1$ are obtained separately from two graphs in Fig. 4 and these results are in agreement with our analytical calculations. The parametrisation of $R$ given by Eq. (9) is very useful for setting up the coupled Maxwell-Lorentz equations, as described in the following section.

\section[\label{sec:level3}]{Coupled Maxwell-Lorentz equations and Gain calculation}
We now build up the analysis of the gain mechanism for a $\check{\text{C}}$erenkov FEL by setting up the Maxwell-Lorentz equations. We would like to emphasize that the approach based on Maxwell-Lorentz equation is familiar for the case of conventional FELs \cite{ColsonFEL1}, BWOs \cite{LevushBWO} and SP-FELs \cite{VinitPRE}. By substituting the value of $R$ in Eq. (4), we obtain the expression for the amplitude of longitudinal electric field as:
\begin{eqnarray}
E_z=\frac{iIZ_0}{2\beta\gamma \Delta y}\bigg(\frac{i\chi}{\mu}e^{-2\Gamma h}+\chi_1e^{-2\Gamma h}-1\bigg)\langle e^{-i\psi}\rangle.~~~~
\end{eqnarray}
The first part of the right hand side of the above expression depends on $\mu$, and is responsible for the growth of the surface electromagnetic mode. Remaining terms are independent of the growth rate parameter and represent the ac space-charge effect in the longitudinal field. This approach of separating the total electromagnetic field into surface mode field and space charge field is similar to the approach described in Ref. \cite{Space}, where it is stated that \textquotedblleft \textit{The total fields from an arbitrary, spatially periodic current are shown to consist of a pole term, which is identified as the structure field, and a remainder, which is identified as the space charge field}\textquotedblright. Dynamics of the electron beam is governed by the surface mode field, as well as the space charge field. We write the space-charge field as $E_{sc}$ and the amplitude of surface-mode as $E$ in the further calculations. We can replace the growth rate parameter by $d/dz$ in Eq. (12), and by including the group velocity $v_g$, we get the following time-dependent differential equation for $E$:
\begin{eqnarray}
\frac{\partial E}{\partial z}+\frac{1}{v_g}\frac{\partial E}{\partial t}=\frac{-IZ_0\chi}{2\beta\gamma \Delta y}e^{-2\Gamma h}\langle e^{-i\psi}\rangle.
\end{eqnarray}    
The space charge field is given by:
\begin{eqnarray}
E_{sc}=\frac{-iIZ_0}{2\beta\gamma \Delta y}(1-\chi_1e^{-2\Gamma h})\langle e^{-i\psi}\rangle.
\end{eqnarray}

Next, we discuss the longitudinal dynamics of the $i$th electron in presence of the surface mode field and the space charge field. We neglect the transverse motion of electron beam and obtain the following equations for the evolution of energy and phase of $i$th electron:
\begin{eqnarray}
\frac{\partial \gamma_i}{\partial z}+\frac{1}{v}\frac{\partial \gamma_i}{\partial t}=\frac{e}{mc^2}(E+E_{sc})e^{i\psi_i}+c.c.,~~~~~~ 
\end{eqnarray}
\begin{eqnarray}
\frac{\partial \psi_i}{\partial z}+\frac{1}{v}\frac{\partial \psi_i}{\partial t} =\frac{\omega}{c\beta^3\gamma^2} \bigg (\frac{\gamma_i-\gamma_p}{\gamma_p}\bigg).~~~~~~~~~~~
\end{eqnarray}
Here, $\gamma_p$=$1/\sqrt{1-v_p^2/c^2}$ is the relativistic Lorentz factor. The subscript $p$ is meant for the resonant particle. At resonance, the electron velocity $v$ is equal to the phase velocity $v_p$ of the co-propagating evanescent surface mode. Equations (13)-(16) can be written in more elegant form by defining the following dimensionless variables:
\begin{equation}
\xi = z/L,
\end{equation}
\begin{eqnarray}
\tau=\bigg( t -\frac{z}{v _p}\bigg) \bigg(\frac{1}{v_g}-\frac{1}{v_p} \bigg)^{-1} \frac{1}{L},
\end{eqnarray}
\begin{equation}
\eta_i=\frac{k_0L}{\beta^2\gamma^3}(\gamma_i-\gamma_p),
\end{equation}
\begin{equation}
\mathcal{E}=\frac{4\pi k_0L^2}{I_AZ_0\beta^2\gamma^3}E,
\end{equation}
\begin{equation}
\mathcal{E}_{sc}=\frac{4\pi k_0L^2}{I_AZ_0\beta^2\gamma^3}E_{sc},
\end{equation}
\begin{equation}
\mathcal{J}=2\pi \frac{\chi}{\Delta y}\frac{I}{I_A}\frac{k_0L^3}{\beta^3\gamma^4}e^{-2\Gamma h}.
\end{equation}
Here, $\xi$ is the dimensionless distance, which varies from 0 to 1, and $\tau$ is the dimensionless time variable, having an offset of $z/v_p$ from the real time $t$. The normalised energy detuning of the $i$th electron is $\eta_i$, $\mathcal{E}$ is the dimensionless surface mode field, and $\mathcal{E}_{sc}$ represents dimensionless space charge field. The dimensionless beam current is written as $\mathcal{J}$ and $I_A$=$4\pi\epsilon_0mc^3/e$=$17.04$ kA is the Alfv$\acute{\text{e}}$n current. With these dimensionless variables, the set of Eqs. (13)-(16) assume the form:
 \begin{equation}
 \frac{\partial \mathcal{E}}{\partial \xi}+\frac{\partial \mathcal{E}}{\partial \tau}= -\mathcal{J}\langle e^{-i\psi}\rangle,~~~~~~
 \end{equation}
 \begin{equation}
 \frac{\partial \eta_i}{\partial \xi}=(\mathcal{E}+\mathcal{E}_{sc})e^{i\psi_i}+c.c.,
 \end{equation}
  \begin{equation}
  \frac{\partial \psi_i}{\partial \xi}=\eta_i,~~~~~~~~
  \end{equation}
 \begin{equation}
 \mathcal{E}_{sc}=i\Theta\langle e^{-i\psi}\rangle,
 \end{equation}
where $\Theta$=$(\mathcal{J}/\chi L)(\chi_1-e^{2\Gamma h})$. These coupled Maxwell-Lorentz equations govern the behaviour of the $\check{\text{C}}$erenkov FEL driven by flat electron beam. We would like to emphasize that our approach and equations described in this section are same as given in Ref. \cite{VinitPRE}, except that the group velocity is positive here, which was negative in Ref. \cite{VinitPRE}. This affects the solution of the equations.

In general, one needs to solve Eqs. (23)-(26) numerically with the given initial conditions for the detailed analysis of the system. However, we can find an analytical solution of these equations in the small-signal, small-gain regime. We will proceed with the time-independent form of Eqs. (23)-(25), and neglect the space charge term. Defining the differential gain as $(1/\mathcal{E})(d\mathcal{E}^2/d\xi)$ and following the procedure closely given in Ref. \cite{Braubook} for conventional undulator based FEL, we get the following expression for the small-signal gain:
 \begin{eqnarray}
 G(\eta_0)=4\mathcal{J}\bigg(\frac{1-\cos\eta_0-\eta_0\sin\eta_0/2}{\eta_0^3} \bigg).
 \end{eqnarray} 
 The term in parentheses is the usual gain function and $\eta_0$=$(k_0L/\beta^2\gamma^3)(\gamma-\gamma_p)$ is the normalised energy detuning at $\xi$=0. Gain function has a maximum value of 6.75$\times$ $10^{-2}$ at $\eta_0$=2.6. By substituting the maximum value of gain function and $\mathcal{J}$ from Eq. (22), we obtain the following expression for small-signal gain in a single pass of CFEL:
 \begin{equation}
 G=4\times 6.75\times 10^{-2}\times 2\pi \frac{\chi}{I_A}\frac{I}{\Delta y}\frac{k_0L^3}{\beta^3\gamma^4}e^{-2\Gamma h} 
 \end{equation} 
Gain increases linearly with the surface current density, and has cubic dependence on length of the dielectric slab. It has an negative exponential dependence on the beam height $h$, and dependences on dielectric constant $\epsilon$ and slab thickness $d$ are given through the parameter $\chi$.

The gain of a CFEL crucially depends upon the diffraction effects in the electromagnetic surface mode. Due to the diffraction, the optical beam size increases, resulting in partial overlap of the optical mode with the electron beam; which reduces the gain of the CFEL. One has to choose the electron beam size $\Delta y$ same as optical beam size for maximum overlapping. The appropriate optical beam size can be estimated by considering the diffraction of electromagnetic fields in $y$ direction \cite{KimPRSTB,VinitPRSTB}. By doing so, we find the effective beam size that needs to be taken in Eq. (28) as $\Delta y$=$\sqrt{\lambda L/2\beta_g}$, where $\beta_g c$ is the group velocity. Note that we have taken $\Delta y$ as $\sqrt{2\pi}$ times the rms beam width. The group velocity can be estimated by evaluating the slope of dispersion curve in Fig. 2. We obtain the group velocity $v_g$=0.236$c$ for the parameters in Table I. The value of small signal gain predicted by our calculations is about 20 $\%$.

We want to emphasize that the expression for gain obtained from our analysis, and the expression derived by Walsh $et~al.$ in the Ref. \cite{Walsh3} give comparable results in the relativistic regime. In Ref. \cite{Walsh3}, the gain analysis has been done for the relativistic regime, while our analysis is applicable to both relativistic, as well as non-relativistic regime. 

We now discuss the calculation of power in the surface mode. Interestingly, this can be done in two different ways. First, the coupled Maxwell-Lorentz equations, which we have derived earlier in this section can be used to evaluate power in the surface mode by using conservation of energy, as discussed in Appendix A. By equating the energy lost by the electron beam to the energy gained by the electromagnetic fields, we obtained the expression for power per unit beam width of the surface mode as:
\begin{equation}
\frac{P}{\Delta y}=\frac{2\beta\gamma}{\chi Z_0}\bigg(\frac{mc^2\beta^2\gamma^3}{ek_0L^2} \bigg)^2e^{2\Gamma h}\vert\mathcal{E}\vert^2.
\end{equation} 
Note that the parameter $\chi$ appears in the above expression. Second, we can explicitly evaluate the power by integrating the Poynting vector in dielectric as well as free space in Fig.~1. This calculation has bee performed in Appendix B, which gives us:
\begin{equation}
\frac{P}{\Delta y}=\frac{\beta \gamma^3}{k_0}\bigg[1+\frac{1}{\epsilon^2a^2}+\frac{k_0d(1+a^2)}{\epsilon \gamma a^2}\bigg]\epsilon_0 cE^2e^{2\Gamma h}.
\end{equation}
Using the expression for $\chi$ given in Eq.~(10) and expression for $\mathcal{E}$ given in Eq.~(20), we find that the expression for power evaluated using the two approaches, which are given by Eqs.~(29) and (30) are exactly identical. This confirms that the formulation of the beam wave interaction in terms of $\chi$ parameter is correct.

\section{\label{sec:level4}Growth rate calculation} 
Analysis in the previous section was done for the small signal, small gain regime. Another regime of interest is the small signal, high gain regime, where we calculate the growth rate in the system. Several authors have presented the calculation of growth rate in CFEL \cite{Brau1,Owens1,Li3}. In this section, we perform the calculation of growth rate in CFEL using collective variables. These variable have been introduced for the study of conventional FELs \cite{Collective} and later extended to study the start-up conditions in SP-FELs \cite{VinitPRE}. For the small signal regime, we assume a perturbative solution of the coupled Maxwell-Lorentz equations. We have neglected the space charge effect here. For simplicity, we assume monoenergetic and unbunched electron beam at the entrance, i.e. $\langle e^{-i\psi_0}\rangle $=0. We can then write the equilibrium solution of Eqs. (23)-(25) as $\mathcal{E}$=0, $\eta_i$=$\eta_0$ and $\psi_i$=$\eta_0\xi+\psi_{i,0}$. We define the perturbative solutions as: $\mathcal{E}_p$=$\mathcal{E}$, $\eta_{i,p}$=$\eta_0+\delta\eta_i$ and $\psi_{i,p}$=$\psi_i+\delta\psi_i$. The collective variables are introduced as:
\begin{equation}
p=\langle \delta\psi e^{-i\psi_0} \rangle,
\end{equation}
\begin{equation}
q=\langle \delta\eta e^{-i\psi_0}\rangle.
\end{equation}
Using above variables, we linearised the set of Eqs. (23)-(25) and by keeping the terms only up to first order, we obtain:
\begin{equation}
\frac{\partial p}{\partial \xi}= q-i\eta_0p,
\end{equation}
\begin{equation}
\frac{\partial q}{\partial \xi}= \mathcal{E}-i\eta_0q,
\end{equation}
\begin{equation}
\frac{\partial \mathcal{E}}{\partial \xi}= i\mathcal{J}p.
\end{equation}
In order to solve the above equations, we assume solution of the type $e^{\nu \xi}$, i.e. $p$=$p_0e^{\nu\xi}$, $q$=$q_0e^{\nu\xi}$ and $\mathcal{E}$=$\mathcal{E}_0e^{\nu\xi}$. With these solutions, Eqs. (33)-(35) now assume the form:
\begin{eqnarray}
\nu p_0=q_0-i\eta_0p_0,~~~~\nu q_0=\mathcal{E}_0-i\eta_0q_0, ~~~\nu\mathcal{E}_0=i\mathcal{J}p_0.
\end{eqnarray}
Above expression can be solved to obtain the following cubic equation in the growth rate parameter:
\begin{eqnarray}
\nu ^3+2i\eta_0\nu^2-\eta_0^2\nu=i\mathcal{J}
\end{eqnarray}
The growth rate will be maximum for $\eta_0$=0. Solving above equation for the positive value of real $\nu$ and substituting $\mathcal{J}$ from Eq. (22), we obtain the maximum growth rate as:
\begin{equation}
\nu_{real}=\frac{\sqrt{3}}{2L}\bigg( 2\pi \frac{\chi}{\Delta y}\frac{I}{I_A}\frac{k_0L^3}{\beta^3\gamma^4}e^{-2\Gamma h} \bigg)^{1/3}.
\end{equation} 
The growth rate depends on cube root of the beam current density. This form of growth rate is already familiar in hydrodynamic approach \cite{Owens1,Brau1,Li3}. Note that we have used flat  electron beam in our calculations. The growth rate for a thick beam having thickness $\Delta x$ in the $x$ direction has been calculated by Li $et~al.$ \cite{Li3}, using hydrodynamic approach. If we take limit $\Delta x$$\rightarrow$0 in the formula given in Ref. \cite{Li3}, we recover Eq. (38).

Using parameters listed in Table 1, we calculated the value of growth rate parameter as 5.2 per m. The growth rate calculated by Andrew and Brau \cite{Brau1} for these parameters, using the three dimensional analysis is about 10 per m. Note that we have used a simple 3D analysis only to estimate effective beam size, while analysis given in Ref. \cite{Brau1} includes three dimensional variation of electromagnetic fields rigorously. However, if we take effective beam size $\Delta y$ as calculated in Ref. \cite{Brau1}, we obtain the value of the growth rate same as given in Ref. \cite{Brau1}. The expected value of growth rate parameter is 250 to 450 per m for the Dartmouth experiment \cite{Owens2}. The value of growth rate parameter obtained from two different analyses are approximately same, but not in agreement with the results of Dartmouth experiment. It is likely that a larger growth rate was measured in Dartmouth experiment due to coherent spontaneous emission.

\section{\label{sec:level5}Numerical simulations}
Next, we discuss about solution of Maxwell-Lorentz equations in the non-linear regime, in order to understand the saturation behaviour of the system. We have written a computer code based on Leapfrog method to solve set of Eqs. (23)-(26). For initial conditions, we assumed a monoenergetic electron beam with $10^5$ particles. The initial electric field is set to be very small. To initialise the electron beam in the phase space, we have used quiet start scheme. In the quiet start scheme, electrons are assumed to have uniform distribution in phase space. The phase of $n$th electron is set to be $2\pi n/N$, where $N$ is total number of particles. This ensure that $\langle e^{-i\psi}\rangle$=0 at $\xi$=0. The total length of system is divided into number of small steps, having step size $\Delta \xi$=0.01.

In the Leapfrog scheme, we require the value of variables at $\xi$=0, and the value of terms on the right side in Eqs.~(23-26) at  $\Delta\xi/2$, to find the value of variables at $\Delta \xi$. In order to evaluate the terms on the right side in Eqs.~(23-26), variables are evaluated at $\Delta\xi/2$ with the help of Eqs. (23)-(25), using Euler method.  Next, the values of variables at $\Delta\xi/2$ are set as initial conditions, and value of variables at $\Delta \xi$ are used in the right hand side of Eqs.~(23-26) to find the solution of Eqs. (23)-(26) at $3\Delta\xi/2$. This scheme is repeated step by step, and we ensure that the energy conservation [Eq. (A1)] is satisfied in each step of integration.

\begin{figure}
\includegraphics[width=17.2cm]{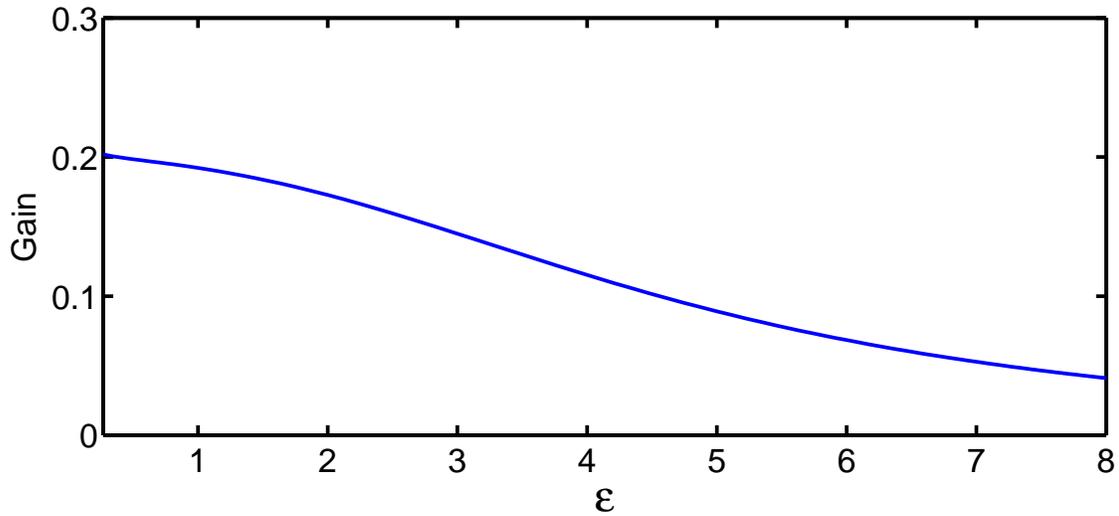}
\caption{Plot of gain as a function of input electric field in a CFEL oscillator.}
\end{figure}
\begin{figure}
\includegraphics[width=17.2cm]{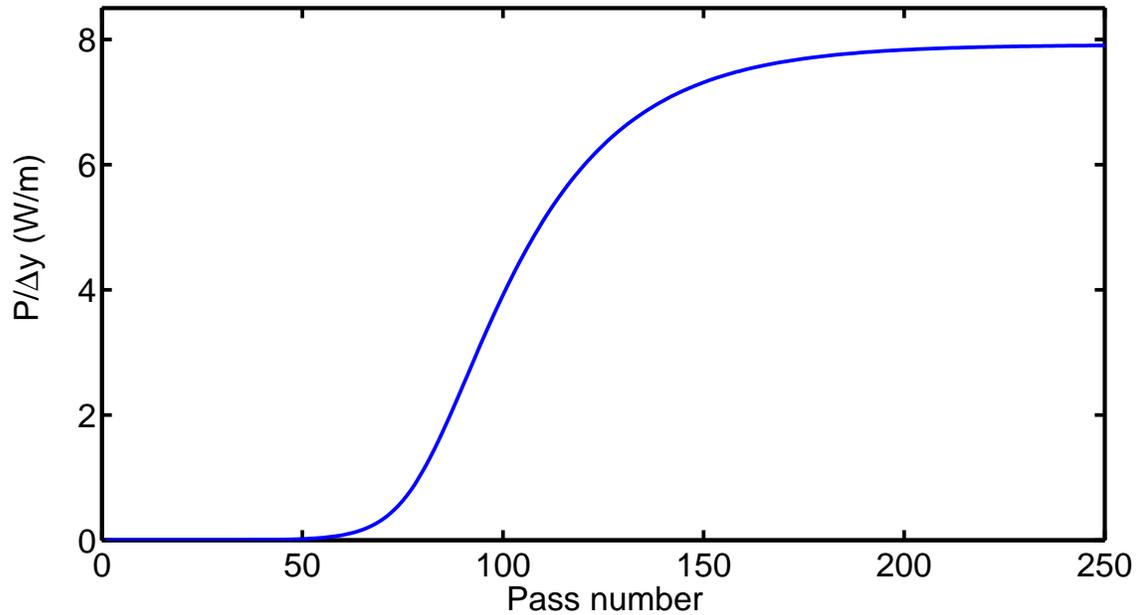}
\caption{Plot of power per unit beam width in the surface mode as a function of number of passes.}
\end{figure}

\begin{figure}
\includegraphics[width=17.2cm]{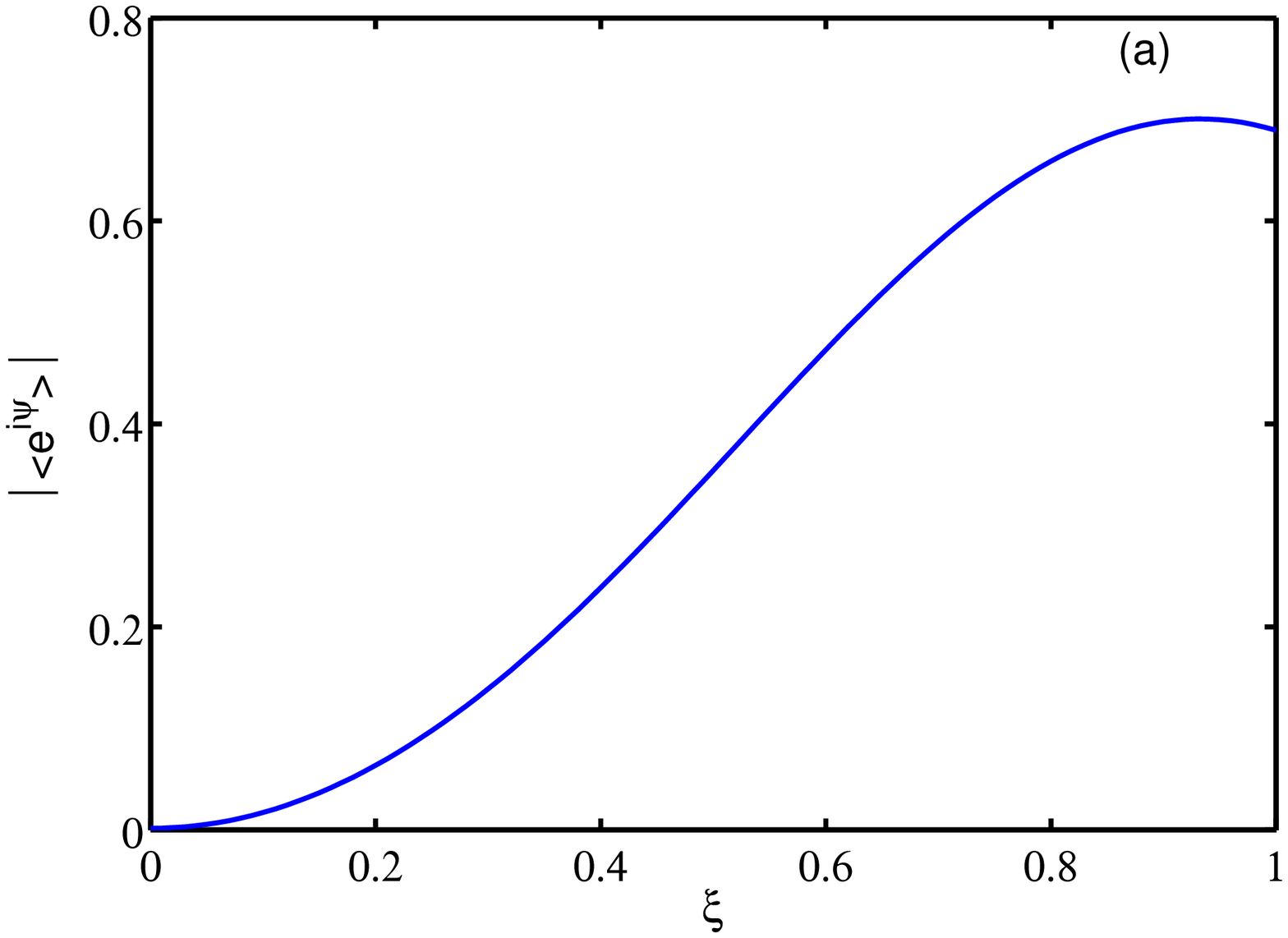}
\includegraphics[width=17.2cm]{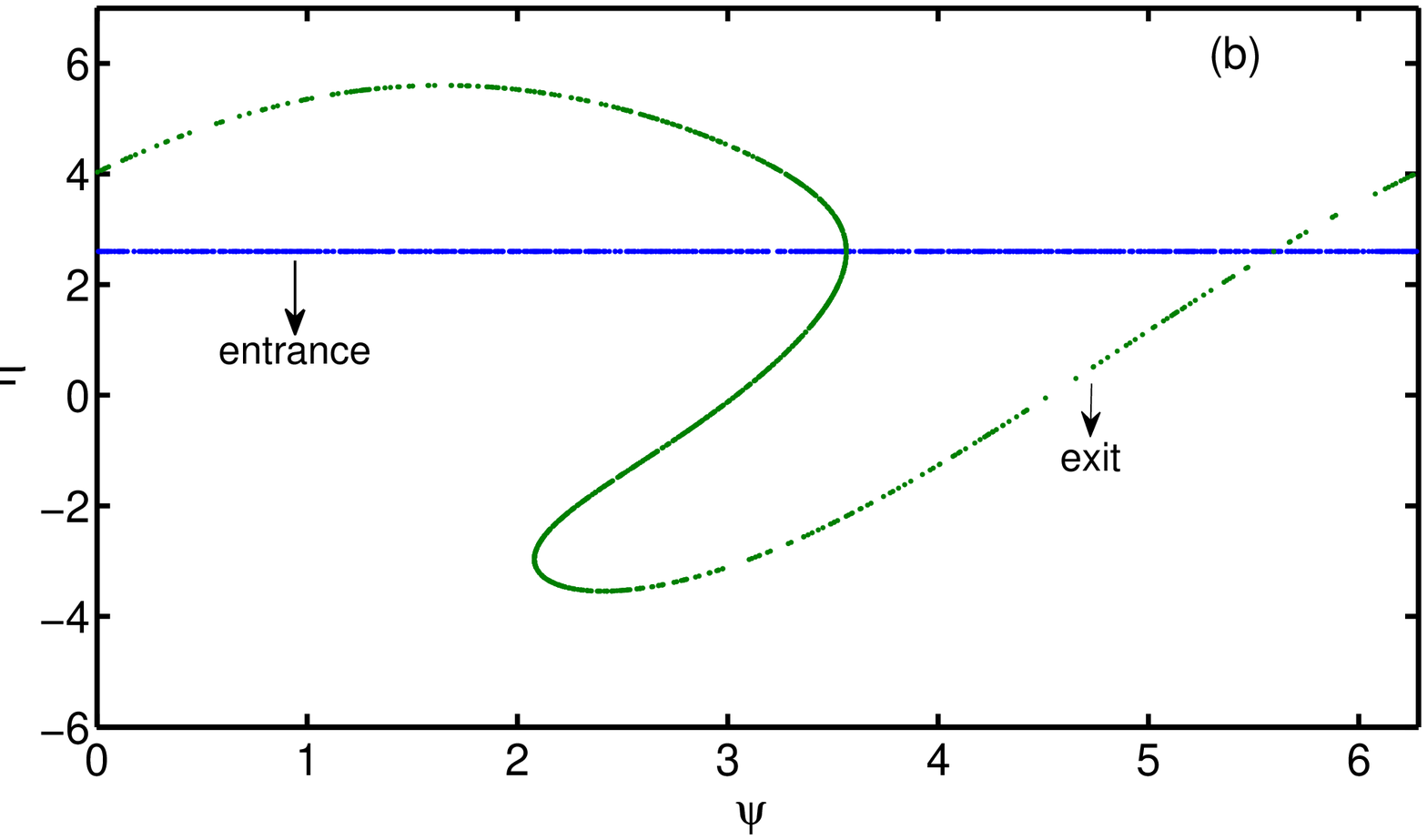}
\caption{(a) Plot of growth of the bunching parameter along the interaction length at saturation of the power in the surface mode. (b) The phase space  of electron beam at the entrance and at the exit of interaction region at saturation.}
\end{figure}

We now discuss the results of our numerical simulations. The parameters used in the code are listed in Table \ref{tab:table1}. Figure 5 shows the gain as a function of initial electric field. We obtain the small-gain of about 20 $\%$. This value of small-signal gain is consistent with our analytical calculations. The gain decreases with the magnitude  of electric field and finally saturates. $\check{\text{C}}$erenkov FEL is a low gain system for the chosen parameters. Hence, in order to get an appreciable output power, we need to operate the device in the oscillator configuration. We propose that a set of mirrors may be used to form a resonator for the oscillator configuration. One mirror is assumed to have 100 $\%$ reflectivity, while the second mirror is having the reflectivity of field amplitude as 98 $\%$. In this configuration, the electromagnetic field reflected at the end of one pass becomes input field for the next pass. 

We examined the non-linear solution of the Maxwell-Lorentz equations by performing numerical simulation in multi-pass operation. Figure~6 shows the growth in the power of the surface mode with the number of passes. The power builds up slowly in this low-gain system, and saturates after 250 passes. The output optical power per unit mode width after saturation is obtained as 7.9 W/m. The input power per unit width of the electron beam is about 1 kW/m. This gives us efficiency of about 0.8 $\%$ at saturation. As discussed by Walsh and Murphy \cite{Walsh2}, the upper bound of the efficiency for power conversion in a CFEL can be written as:
\begin{equation}
\eta_{eff}=\frac{\beta^3\gamma^3}{(\gamma-1)}\frac{\lambda}{L}
\end{equation}
We get an upper bound of 1.4 $\%$ for the efficiency of a $\check{\text{C}}$erenkov FEL. This is in well agreement with the results of numerical simulations as the analytic expression is only a rough estimate for the maximum value of efficiency.

We also examined the evaluation in the phase space distribution of the electrons along the interaction region. Figure 7(a) shows amplitude of bunching parameter $\vert\langle e^{-i\psi}\rangle \vert$ along the distance after the saturation of output power. We observed that the electrons are nicely bunched at the exit of the interaction region and the amplitude of bunching parameter is about 0.7. Similar mechanism is observed in the plot of phase space distribution of electrons at the entrance and at the exit of the interaction region. We clearly see in the Fig. 7(b) that the electrons are randomly distributed at the entrance, and become bunched at the end due to the interaction with the co-propagating surface mode.

\section{\label{sec:level6}Discussion and conclusion}
In this paper, we have presented an analysis of $\check{\text{C}}$erenkov FEL driven by a flat electron beam for the single slab geometry, by setting up Maxwell-Lorentz equations. For conventional undulator based FELs, the approach based on Maxwell-Lorentz equations has been extremely successful, particularly to understand the behaviour in the non-linear regime, and to incorporate realistic effects. However, most of the analyses of CFEL in slab geometry so far have used either the  Mazwell-Vlasov equation approach, or the hydrodynamic approach, which are useful in the linear regime. There has been earlier attempt to analyze the CFEL system for slab geometry based on Maxwell-Lorentz equations in Refs. \cite{Asgek1,Asgekar2,Asgekar3}. However, the evanescent nature of the surface mode has not been included in these analyses in a rigorous way, while setting up the coupled Maxwell-Lorentz equations. As a result, these analyses do not describe the dependency on the height of electron beam from the dielectric surface properly, and also do   not describe the case of flat beam appropriately since the volume current density becomes very large. In our analysis, we have included the evanescent nature of the surface mode in a rigorous way, and also included the effect of space charge by taking into account the field due to incident and reflected evanescent wave.

In our analysis, we have not considered the effects due to energy spread, beam emittance and three dimensional variations of the optical field. Since the Maxwell-Lorentz based approach that we have developed in this paper is identical to the approach developed for SPFEL \cite{VinitPRE,KimPRSTB,VinitPRSTB} and conventional FEL \cite{Collective}, we can add these effects by similar way, which we propose to do in the future.

Our analysis is build up on the earlier analysis of Smith-Purcell FEL, where the parameters $\chi$ and $\chi_1$ obtained from the Laurent expansion of reflectivity of the system around the singularity were used for setting up of Maxwell-Lorentz equations. In the analysis presented in Ref.~\cite{VinitPRE}, the expression for power in the surface mode was obtained by applying energy conservation in the coupled Maxwell-Lorentz equations. For the case of SP-FEL, it is not possible to derive a simple expression for power in the surface mode by integration of Poynting vector since reflection grating is a complex electromagnetic system. Thus, the power calculation in terms of $\chi$ parameter derived using energy conservation approach could not be cross checked with the expression derived by integration of Poynting vector. On the other hand, for the analysis of CFEL system presented in this paper, it has been possible to derive an analytical expression for the power by integrating the Poynting vector since dielectric is a much simpler system compared to a reflection grating, which has infinite number of space harmonics in the expression for electromagnetic field. We have thus been able to check that for the general framework of setting up of Maxwell-Lorentz equations in terms of the $\chi$ parameter, the expression for power flow is calculated correctly. Another interesting observation is that unlike SP-FEL system, here we have been able to derive analytical expressions for $\chi$ and $\chi_1$, which can be calculated only numerically for the SP-FEL case.

We have considered the parameters used in the Dartmouth experiment \cite{Owens2} to perform calculations. The output power reported in the Dartmouth experiment \cite{Owens2} was of the order of picowatt and growth rate was predicted around 250-450 per m. To get an appreciable output power, authors in Ref. \cite{Owens2} suggested the use of the flat electron beam to drive the CFEL. We performed the analysis with flat electron beam and obtained the output power of around 240 mW with an efficiency of about 0.8 $\%$ at saturation in the oscillator configuration. Outcoupling of  THz radiation can be done by putting a hole in the outcoupling mirror and the radiation power can be directed to the useful experiments. In this way, $\check{\text{C}}$erenkov FELs can fulfil the requirements of several industrial and scientific applications which require copious coherent THz radiation.

We would like to mention that the formula for $\chi$ and $\chi_1$ that we have described in this paper can be used for the case of negative index material also, and our analysis can be extended to understand the behaviour of CFEL in BWO configuration which is described in Ref. \cite{Li3}. Also, the analysis can be extended for the case of finite thickness of the electron beam in the $x$-direction. This can be done by treating the beam as combination of continuous layers of flat beam.

To conclude, we have presented an analysis for the working of CFEL by studying the singularity in the reflectivity of the dielectric slab. We have set up the coupled Maxwell-Lorentz equations, taking into the evanescent nature of surface mode and also the space charge field. This approach is suitable for writing computer program for analysis of the CFEL system taking realistic effects such as energy spread, beam emittance, etc. into account, and also to analyze the nonlinear behaviour of the system. For conventional FEL \cite{ColsonFEL1} and SPFEL \cite{VinitPRE,KimPRSTB,VinitPRSTB,VinitFEL05}, this approach has already been very successful, and by extending this approach to $\check{\text{C}}$erenkov FELs, we have stepped forward towards having a unified theory for all FELs.
Our analysis can be useful for the design and the operation of a compact $\check{\text{C}}$erenkov FEL working in the THz regime.

\appendix

\section{ENERGY CONSERVATION}
In this appendix, we derive the expression for power in the surface mode by using the energy conservation in the coupled Maxwell-Lorentz equations. This approach has been discussed in detail for the SP-FELs \cite{VinitPRE}. Here, we are summarizing the results for $\check{\text{C}}$erenkov FELs. In steady state, the set of Eqs. (23)-(26) can be combined to obtain the following expression:
\begin{equation}
\frac{\partial}{\partial\xi}\big(\vert \mathcal{E}\vert^2 + \mathcal{J}\langle \eta \rangle \big)=0.
\end{equation} 
It indicates that the energy lost by the electron beam while travelling down the interaction region, appears in the surface modes. This is the principle of conservation of energy. Using Eq. (A1), we can express the energy lost per unit time by the electron beam and equate it to the power developed in the surface mode. By doing the required algebra in this way, we obtain the following expression of the power in the surface mode:
\begin{equation}
\frac{P}{\Delta y}=\frac{2\beta\gamma}{\chi Z_0}\bigg(\frac{mc^2\beta^2\gamma^3}{ek_0L^2} \bigg)^2e^{2\Gamma h}\vert\mathcal{E}\vert^2.
\end{equation}
This expression is used to calculate the power in our numerical simulations.

\section{CALCULTION OF THE GROUP VELOCITY AND POWER IN THE SURFACE MODE}
Here, we calculate group velocity and power flow in the surface electromagnetic mode supported by the configuration, described earlier in Sec. I, using Poynting vector. The basic geometry of a CFEL, consists of a dielectric slab resting on an ideal conductor. The electromagnetic fields for this structure have been calculated by several authors \cite{Walsh2,Asgek1}. We assign region I to vacuum and region II to the dielectric slab as shown in Fig. 1. The electromagnetic fields in the region I can be written as:
\begin{eqnarray}
H_y^I(x,z,t)=H\exp[i\psi-\Gamma(x+h)]+c.c., ~~~~~~~~~~~~~~\\
E_x^I(x,z,t)=(H/\beta\epsilon_0c)\exp[i\psi-\Gamma(x+h)]+c.c.,~~~~~\\
E_z^I(x,z,t)=(-iH/\beta\gamma \epsilon_0c)\exp[i\psi-\Gamma(x+h)]+c.c..
\end{eqnarray}
Here, $\psi$=$k_0z-\omega t$ and $H$ is strength of the magnetic field at the dielectric surface. In region II, we find electromagnetic fields as:
\begin{eqnarray}
H_y^{II}(x,z,t)=\frac{ \epsilon\Gamma}{k_1}\frac{\cos[k_1(x+h+d)]}{\sin(k_1d)}H\exp(i\psi)+c.c.,~~~
\end{eqnarray}
\begin{eqnarray}
E_x^{II}(x,z,t)=\frac{k_0\Gamma}{\omega \epsilon_0k_1}\frac{\cos[k_1(x+h+d)]}{\sin(k_1d)}H\exp(i\psi)+c.c.,
\end{eqnarray}
\begin{eqnarray}
E_z^{II}(x,z,t)=\frac{-i\Gamma}{\omega \epsilon_0}\frac{\sin[k_1(x+h+d)]}{\sin(k_1d)}H\exp(i\psi)+c.c.,~~
\end{eqnarray}
where $k_1=k_0\sqrt{\epsilon\beta^2-1}$. Power flow in the electromagnetic fields can be calculated by integrating the Poynting vector over the transverse area. Total power in the surface mode is sum of the power in region I and in region II. We obtain the following expression for power in the surface mode:
\begin{equation}
\frac{P}{\Delta y}=\frac{\beta \gamma^3}{k_0}\bigg[1+\frac{1}{\epsilon^2a^2}+\frac{k_0d(1+a^2)}{\epsilon \gamma a^2}\bigg]\epsilon_0 cE^2e^{2\Gamma h}.
\end{equation} 

Next, we drive the expression for the total energy stored in the electromagnetic fields. The energy stored in the fields can be evaluated by integrating the energy density $\mathcal{U}$=$(\epsilon E^2 +\mu H^2)/2$ over the volume of the required region. Total energy stored in electromagnetic fields is sum of energy in vacuum and in the dielectric medium, which we obtain as:
\begin{eqnarray}
\frac{U}{\Delta y}=\frac{\gamma^3L}{k_0c}\bigg[1+\frac{1}{\epsilon^2a^2}+\frac{k_0d\beta^2(1+a^2)}{\gamma a^2}\bigg]\epsilon_0 cE^2e^{2\Gamma h}.
\end{eqnarray}
By knowing the expression for the power $P$ and the energy stored $U$, one can find the energy velocity as $v_e$=$PL/U$. Using Eq. (B7) and the Eq. (B8), we obtain the following analytical expression for the energy velocity:
\begin{equation}
v_e=\frac{v_p\big[\beta^2\gamma^3(\epsilon-1)+k_0d\epsilon(1+a^2)\big]}{\big[\beta^2\gamma^3(\epsilon-1)+k_0d\epsilon^2\beta^2(1+a^2)\big]}~,
\end{equation} 
where $v_p$ is the phase velocity of the surface mode. For each value of $\beta$, we find $v_e$ from Eq. (B9) and group velocity from the dispersion curve in Fig. 2. We find that for $\check{\text{C}}$erenkov FEL, the energy velocity is equal to the group velocity. 

\section*{Acknowledgment}
One of us (YK) gratefully acknowledges Homi Bhabha National Institute, Department of Atomic Energy (India) for financial support.


\begin{thebibliography}{42}
\expandafter\ifx\csname natexlab\endcsname\relax\def\natexlab#1{#1}\fi
\expandafter\ifx\csname bibnamefont\endcsname\relax
  \def\bibnamefont#1{#1}\fi
\expandafter\ifx\csname bibfnamefont\endcsname\relax
  \def\bibfnamefont#1{#1}\fi
\expandafter\ifx\csname citenamefont\endcsname\relax
  \def\citenamefont#1{#1}\fi
\expandafter\ifx\csname url\endcsname\relax
  \def\url#1{\texttt{#1}}\fi
\expandafter\ifx\csname urlprefix\endcsname\relax\def\urlprefix{URL }\fi
\providecommand{\bibinfo}[2]{#2}
\providecommand{\eprint}[2][]{\url{#2}}

\bibitem[{\citenamefont{$\check{\text{C}}$erenkov}(1937)}]{Cerenkov}
\bibinfo{author}{\bibfnamefont{P.~A.} \bibnamefont{$\check{\text{C}}$erenkov}},
  \bibinfo{journal}{Phys. Rev.} \textbf{\bibinfo{volume}{52}},
  \bibinfo{pages}{378} (\bibinfo{year}{1937}).

\bibitem[{\citenamefont{Ginzburg}(1947)}]{Ginzburg}
\bibinfo{author}{\bibfnamefont{V.}~\bibnamefont{Ginzburg}},
  \bibinfo{journal}{Dokl. Akad. Nauk SSSR} \textbf{\bibinfo{volume}{56}},
  \bibinfo{pages}{253} (\bibinfo{year}{1947}).

\bibitem[{\citenamefont{Walsh and Murphy}(1982)}]{Walsh2}
\bibinfo{author}{\bibfnamefont{J.~E.} \bibnamefont{Walsh}} \bibnamefont{and}
  \bibinfo{author}{\bibfnamefont{J.~B.} \bibnamefont{Murphy}},
  \bibinfo{journal}{IEEE J. Quantum Electron} \textbf{\bibinfo{volume}{QE-18}},
  \bibinfo{pages}{1259} (\bibinfo{year}{1982}).

\bibitem[{\citenamefont{Walsh et~al.}(1984)\citenamefont{Walsh, Johnson,
  Dattoli, and Renieri}}]{Walsh3}
\bibinfo{author}{\bibfnamefont{J.}~\bibnamefont{Walsh}},
  \bibinfo{author}{\bibfnamefont{B.}~\bibnamefont{Johnson}},
  \bibinfo{author}{\bibfnamefont{G.}~\bibnamefont{Dattoli}}, \bibnamefont{and}
  \bibinfo{author}{\bibfnamefont{A.}~\bibnamefont{Renieri}},
  \bibinfo{journal}{Phys. Rev. Lett.} \textbf{\bibinfo{volume}{53}},
  \bibinfo{pages}{779} (\bibinfo{year}{1984}).

\bibitem[{\citenamefont{Garate et~al.}(1987)\citenamefont{Garate, Shaughnessy,
  and Walsh}}]{Garate2}
\bibinfo{author}{\bibfnamefont{E.~P.} \bibnamefont{Garate}},
  \bibinfo{author}{\bibfnamefont{C.~H.} \bibnamefont{Shaughnessy}},
  \bibnamefont{and} \bibinfo{author}{\bibfnamefont{J.}~\bibnamefont{Walsh}},
  \bibinfo{journal}{IEEE J. Quantum Electron.}
  \textbf{\bibinfo{volume}{QE-23}}, \bibinfo{pages}{1627}
  (\bibinfo{year}{1987}).

\bibitem[{\citenamefont{Tripathi}(1984)}]{Tripathi1}
\bibinfo{author}{\bibfnamefont{V.~K.} \bibnamefont{Tripathi}},
  \bibinfo{journal}{J. Appl. Phys.} \textbf{\bibinfo{volume}{56}},
  \bibinfo{pages}{1953} (\bibinfo{year}{1984}).

\bibitem[{\citenamefont{Sharma and Bhasin}(2007)}]{Bhasin}
\bibinfo{author}{\bibfnamefont{S.~C.} \bibnamefont{Sharma}} \bibnamefont{and}
  \bibinfo{author}{\bibfnamefont{A.}~\bibnamefont{Bhasin}},
  \bibinfo{journal}{Phys. Plasmas} \textbf{\bibinfo{volume}{053101}},
  \bibinfo{pages}{14} (\bibinfo{year}{2007}).

\bibitem[{\citenamefont{Fares and Yamada}(2011)}]{Fares1}
\bibinfo{author}{\bibfnamefont{H.}~\bibnamefont{Fares}} \bibnamefont{and}
  \bibinfo{author}{\bibfnamefont{M.}~\bibnamefont{Yamada}},
  \bibinfo{journal}{Phys. Plasmas} \textbf{\bibinfo{volume}{18}},
  \bibinfo{pages}{093106} (\bibinfo{year}{2011}).

\bibitem[{\citenamefont{Fares}(2012)}]{Fares2}
\bibinfo{author}{\bibfnamefont{H.}~\bibnamefont{Fares}},
  \bibinfo{journal}{Phys. Plasmas} \textbf{\bibinfo{volume}{19}},
  \bibinfo{pages}{053109} (\bibinfo{year}{2012}).

\bibitem[{\citenamefont{Kheiri and Esmaeilzadeh}(2013)}]{Kheiri}
\bibinfo{author}{\bibfnamefont{G.}~\bibnamefont{Kheiri}} \bibnamefont{and}
  \bibinfo{author}{\bibfnamefont{M.}~\bibnamefont{Esmaeilzadeh}},
  \bibinfo{journal}{Phys. Plasmas} \textbf{\bibinfo{volume}{20}},
  \bibinfo{pages}{123107} (\bibinfo{year}{2013}).

\bibitem[{\citenamefont{Freund and Ganguly}(1990)}]{Ganguly}
\bibinfo{author}{\bibfnamefont{H.}~\bibnamefont{Freund}} \bibnamefont{and}
  \bibinfo{author}{\bibfnamefont{A.}~\bibnamefont{Ganguly}},
  \bibinfo{journal}{Nucl. Instrum. Methods Phys. Res. A}
  \textbf{\bibinfo{volume}{296}}, \bibinfo{pages}{462} (\bibinfo{year}{1990}).

\bibitem[{\citenamefont{Freund}(1990)}]{Freund}
\bibinfo{author}{\bibfnamefont{H.~P.} \bibnamefont{Freund}},
  \bibinfo{journal}{Phys. Rev. Lett.} \textbf{\bibinfo{volume}{65}},
  \bibinfo{pages}{2993} (\bibinfo{year}{1990}).

\bibitem[{\citenamefont{de~la Fuente et~al.}(2007)\citenamefont{de~la Fuente,
  van~der Slot, and Boller}}]{fuente1}
\bibinfo{author}{\bibfnamefont{I.}~\bibnamefont{de~la Fuente}},
  \bibinfo{author}{\bibfnamefont{P.~J.~M.} \bibnamefont{van~der Slot}},
  \bibnamefont{and} \bibinfo{author}{\bibfnamefont{K.-J.}
  \bibnamefont{Boller}}, \bibinfo{journal}{Phys. Rev. ST Accel. Beams}
  \textbf{\bibinfo{volume}{10}}, \bibinfo{pages}{020702} (\bibinfo{year}{2007}).

\bibitem[{\citenamefont{de~la Fuente}(2007)}]{Fuentethesis}
\bibinfo{author}{\bibfnamefont{I.}~\bibnamefont{de~la Fuente}}, Ph.D. thesis,
  \bibinfo{school}{Laser Physics and Non-Linear Optics Group, University of
  Twente} (\bibinfo{year}{2007}).

\bibitem[{\citenamefont{Walsh et~al.}(1977)\citenamefont{Walsh, Marshall, and
  Schlesinger}}]{Walsh1}
\bibinfo{author}{\bibfnamefont{J.~E.} \bibnamefont{Walsh}},
  \bibinfo{author}{\bibfnamefont{T.~C.} \bibnamefont{Marshall}},
  \bibnamefont{and} \bibinfo{author}{\bibfnamefont{S.~P.}
  \bibnamefont{Schlesinger}}, \bibinfo{journal}{Phys. Fluids}
  \textbf{\bibinfo{volume}{20}}, \bibinfo{pages}{709} (\bibinfo{year}{1977}).

\bibitem[{\citenamefont{Owens and Brownell}(2003)}]{Owens1}
\bibinfo{author}{\bibfnamefont{I.~J.} \bibnamefont{Owens}} \bibnamefont{and}
  \bibinfo{author}{\bibfnamefont{J.~H.} \bibnamefont{Brownell}},
  \bibinfo{journal}{Phys. Rev. E} \textbf{\bibinfo{volume}{67}},
  \bibinfo{pages}{036611} (\bibinfo{year}{2003}).

\bibitem[{\citenamefont{Andrews and Brau}(2007)}]{Brau1}
\bibinfo{author}{\bibfnamefont{H.~L.} \bibnamefont{Andrews}} \bibnamefont{and}
  \bibinfo{author}{\bibfnamefont{C.~A.} \bibnamefont{Brau}},
  \bibinfo{journal}{J. Appl. Phys.} \textbf{\bibinfo{volume}{101}},
  \bibinfo{pages}{104904} (\bibinfo{year}{2007}).

\bibitem[{\citenamefont{Li et~al.}(2009)\citenamefont{Li, Huo, Imasaki, and
  M.Asakawa}}]{Li1}
\bibinfo{author}{\bibfnamefont{D.}~\bibnamefont{Li}},
  \bibinfo{author}{\bibfnamefont{G.}~\bibnamefont{Huo}},
  \bibinfo{author}{\bibfnamefont{K.}~\bibnamefont{Imasaki}}, \bibnamefont{and}
  \bibinfo{author}{\bibnamefont{M.Asakawa}}, \bibinfo{journal}{Nucl. Instrum.
  Methods Phys. Res. A} \textbf{\bibinfo{volume}{606}}, \bibinfo{pages}{689}
  (\bibinfo{year}{2009}).

\bibitem[{\citenamefont{Sharma and Mishra}(2012)}]{Sharma}
\bibinfo{author}{\bibfnamefont{G.}~\bibnamefont{Sharma}} \bibnamefont{and}
  \bibinfo{author}{\bibfnamefont{G.}~\bibnamefont{Mishra}},
  \bibinfo{journal}{Nucl. Instrum. Methods Phys. Res. A}
  \textbf{\bibinfo{volume}{685}}, \bibinfo{pages}{35} (\bibinfo{year}{2012}).

\bibitem[{\citenamefont{Li et~al.}(2014)\citenamefont{Li, Wang, Hangyo, Wei,
  Yang, and Miyamoto}}]{Li3}
\bibinfo{author}{\bibfnamefont{D.}~\bibnamefont{Li}},
  \bibinfo{author}{\bibfnamefont{Y.}~\bibnamefont{Wang}},
  \bibinfo{author}{\bibfnamefont{M.}~\bibnamefont{Hangyo}},
  \bibinfo{author}{\bibfnamefont{Y.}~\bibnamefont{Wei}},
  \bibinfo{author}{\bibfnamefont{Z.}~\bibnamefont{Yang}}, \bibnamefont{and}
  \bibinfo{author}{\bibfnamefont{S.}~\bibnamefont{Miyamoto}},
  \bibinfo{journal}{Appl. Phys. Lett.} \textbf{\bibinfo{volume}{104}},
  \bibinfo{pages}{194102} (\bibinfo{year}{2014}).

\bibitem[{\citenamefont{Gore et~al.}(1996)\citenamefont{Gore, Asgekar, and
  Sen}}]{Asgek1}
\bibinfo{author}{\bibfnamefont{B.~W.} \bibnamefont{Gore}},
  \bibinfo{author}{\bibfnamefont{V.~B.} \bibnamefont{Asgekar}},
  \bibnamefont{and} \bibinfo{author}{\bibfnamefont{A.}~\bibnamefont{Sen}},
  \bibinfo{journal}{Phys. Scripta.} \textbf{\bibinfo{volume}{53}},
  \bibinfo{pages}{62} (\bibinfo{year}{1996}).

\bibitem[{\citenamefont{Asgekar and Dattoli}(2002)}]{Asgekar2}
\bibinfo{author}{\bibfnamefont{V.~B.} \bibnamefont{Asgekar}} \bibnamefont{and}
  \bibinfo{author}{\bibfnamefont{G.}~\bibnamefont{Dattoli}},
  \bibinfo{journal}{Opt. Commun.} \textbf{\bibinfo{volume}{206}},
  \bibinfo{pages}{373} (\bibinfo{year}{2002}).

\bibitem[{\citenamefont{Asgekar and Dattoli}(2005)}]{Asgekar3}
\bibinfo{author}{\bibfnamefont{V.~B.} \bibnamefont{Asgekar}} \bibnamefont{and}
  \bibinfo{author}{\bibfnamefont{G.}~\bibnamefont{Dattoli}},
  \bibinfo{journal}{Opt. Commun.} \textbf{\bibinfo{volume}{255}},
  \bibinfo{pages}{309} (\bibinfo{year}{2005}).

\bibitem[{\citenamefont{Danos et~al.}(1953)\citenamefont{Danos, Geschwind,
  Lashinsky, and Trier}}]{Danos1}
\bibinfo{author}{\bibfnamefont{M.}~\bibnamefont{Danos}},
  \bibinfo{author}{\bibfnamefont{S.}~\bibnamefont{Geschwind}},
  \bibinfo{author}{\bibfnamefont{H.}~\bibnamefont{Lashinsky}},
  \bibnamefont{and} \bibinfo{author}{\bibfnamefont{A.~V.} \bibnamefont{Trier}},
  \bibinfo{journal}{Phys. Rev.} \textbf{\bibinfo{volume}{92}},
  \bibinfo{pages}{828} (\bibinfo{year}{1953}).

\bibitem[{\citenamefont{Seo. et~al.}(2000)\citenamefont{Seo., Choi, and
  Cho}}]{Seo}
\bibinfo{author}{\bibfnamefont{Y.}~\bibnamefont{Seo.}},
  \bibinfo{author}{\bibfnamefont{E.~H.} \bibnamefont{Choi}}, \bibnamefont{and}
  \bibinfo{author}{\bibfnamefont{G.~S.} \bibnamefont{Cho}},
  \bibinfo{journal}{J. Phys. D: Appl. Phys.} \textbf{\bibinfo{volume}{33}},
  \bibinfo{pages}{654} (\bibinfo{year}{2000}).

\bibitem[{\citenamefont{Owens and Brownell}(2005)}]{Owens2}
\bibinfo{author}{\bibfnamefont{I.~J.} \bibnamefont{Owens}} \bibnamefont{and}
  \bibinfo{author}{\bibfnamefont{J.~H.} \bibnamefont{Brownell}},
  \bibinfo{journal}{J. Appl. Phys.} \textbf{\bibinfo{volume}{97}},
  \bibinfo{pages}{104915} (\bibinfo{year}{2005}).

\bibitem[{\citenamefont{Li et~al.}(2010)\citenamefont{Li, Huo, Imasaki,
  M.Asakawa, and Tsunawaki}}]{Li2}
\bibinfo{author}{\bibfnamefont{D.}~\bibnamefont{Li}},
  \bibinfo{author}{\bibfnamefont{G.}~\bibnamefont{Huo}},
  \bibinfo{author}{\bibfnamefont{K.}~\bibnamefont{Imasaki}},
  \bibinfo{author}{\bibnamefont{M.Asakawa}}, \bibnamefont{and}
  \bibinfo{author}{\bibfnamefont{Y.}~\bibnamefont{Tsunawaki}},
  \bibinfo{journal}{Infrared Phys. Techn.} \textbf{\bibinfo{volume}{53}},
  \bibinfo{pages}{204} (\bibinfo{year}{2010}).

\bibitem[{\citenamefont{Garate et~al.}(1986)\citenamefont{Garate, Moustaizis,
  Buzzi, Rouille, Lamain, Walsh, and Johnson}}]{Garate1}
\bibinfo{author}{\bibfnamefont{E.~P.} \bibnamefont{Garate}},
  \bibinfo{author}{\bibfnamefont{S.}~\bibnamefont{Moustaizis}},
  \bibinfo{author}{\bibfnamefont{J.~M.} \bibnamefont{Buzzi}},
  \bibinfo{author}{\bibfnamefont{C.}~\bibnamefont{Rouille}},
  \bibinfo{author}{\bibfnamefont{H.}~\bibnamefont{Lamain}},
  \bibinfo{author}{\bibfnamefont{J.}~\bibnamefont{Walsh}}, \bibnamefont{and}
  \bibinfo{author}{\bibfnamefont{B.}~\bibnamefont{Johnson}},
  \bibinfo{journal}{Appl. Phys. Lett.} \textbf{\bibinfo{volume}{48}},
  \bibinfo{pages}{1326} (\bibinfo{year}{1986}).

\bibitem[{\citenamefont{Seigel}(2002)}]{Seigel}
\bibinfo{author}{\bibfnamefont{P.~H.} \bibnamefont{Seigel}},
  \bibinfo{journal}{IEEE Trans. Microwave Theory Tech.}
  \textbf{\bibinfo{volume}{50}}, \bibinfo{pages}{911} (\bibinfo{year}{2002}).

\bibitem[{\citenamefont{Clery}(2002)}]{Clery}
\bibinfo{author}{\bibfnamefont{D.}~\bibnamefont{Clery}},
  \bibinfo{journal}{Science} \textbf{\bibinfo{volume}{297}},
  \bibinfo{pages}{761} (\bibinfo{year}{2002}).

\bibitem[{\citenamefont{Carr et~al.}(2002)\citenamefont{Carr, Martin, McKinney,
  K.~Jordan, and Williams}}]{THz}
\bibinfo{author}{\bibfnamefont{G.~L.} \bibnamefont{Carr}},
  \bibinfo{author}{\bibfnamefont{M.~C.} \bibnamefont{Martin}},
  \bibinfo{author}{\bibfnamefont{W.~R.} \bibnamefont{McKinney}},
  \bibinfo{author}{\bibfnamefont{G.~R.~N.} \bibnamefont{K.~Jordan}},
  \bibnamefont{and} \bibinfo{author}{\bibfnamefont{G.~P.}
  \bibnamefont{Williams}}, \bibinfo{journal}{Nature(London)}
  \textbf{\bibinfo{volume}{420}}, \bibinfo{pages}{153} (\bibinfo{year}{2002}).

\bibitem[{\citenamefont{di~Francia}(1960)}]{Toraldo}
\bibinfo{author}{\bibfnamefont{G.~T.} \bibnamefont{di~Francia}},
  \bibinfo{journal}{Nuovo Cimento} \textbf{\bibinfo{volume}{16}},
  \bibinfo{pages}{61} (\bibinfo{year}{1960}).

\bibitem[{\citenamefont{Schachter and Ron}(1989)}]{Levi}
\bibinfo{author}{\bibfnamefont{L.}~\bibnamefont{Schachter}} \bibnamefont{and}
  \bibinfo{author}{\bibfnamefont{A.}~\bibnamefont{Ron}},
  \bibinfo{journal}{Phys. Rev. A} \textbf{\bibinfo{volume}{40}},
  \bibinfo{pages}{876} (\bibinfo{year}{1989}).

\bibitem[{\citenamefont{Kumar and Kim}(2006)}]{VinitPRE}
\bibinfo{author}{\bibfnamefont{V.}~\bibnamefont{Kumar}} \bibnamefont{and}
  \bibinfo{author}{\bibfnamefont{K.-J.} \bibnamefont{Kim}},
  \bibinfo{journal}{Phys. Rev. E} \textbf{\bibinfo{volume}{73}},
  \bibinfo{pages}{026501} (\bibinfo{year}{2006}).

\bibitem[{\citenamefont{Kim and Kumar}(2007)}]{KimPRSTB}
\bibinfo{author}{\bibfnamefont{K.-J.} \bibnamefont{Kim}} \bibnamefont{and}
  \bibinfo{author}{\bibfnamefont{V.}~\bibnamefont{Kumar}},
  \bibinfo{journal}{Phys. Rev. ST Accel. Beams} \textbf{\bibinfo{volume}{10}},
  \bibinfo{pages}{080702} (\bibinfo{year}{2007}).

\bibitem[{\citenamefont{Kumar and Kim}(2009)}]{VinitPRSTB}
\bibinfo{author}{\bibfnamefont{V.}~\bibnamefont{Kumar}} \bibnamefont{and}
  \bibinfo{author}{\bibfnamefont{K.-J.} \bibnamefont{Kim}},
  \bibinfo{journal}{Phys. Rev. ST Accel. Beams} \textbf{\bibinfo{volume}{12}},
  \bibinfo{pages}{070703} (\bibinfo{year}{2009}).

\bibitem[{\citenamefont{Kumar and Kim}(2005)}]{VinitFEL05}
\bibinfo{author}{\bibfnamefont{V.}~\bibnamefont{Kumar}} \bibnamefont{and}
  \bibinfo{author}{\bibfnamefont{K.-J.} \bibnamefont{Kim}}, in
  \emph{\bibinfo{booktitle}{Proceedings of FEL05}} (\bibinfo{year}{2005}), pp.
  \bibinfo{pages}{274--277}.

\bibitem[{\citenamefont{Colson}(1976)}]{ColsonFEL1}
\bibinfo{author}{\bibfnamefont{W.~B.} \bibnamefont{Colson}},
  \bibinfo{journal}{Phys. Lett.} \textbf{\bibinfo{volume}{59A}},
  \bibinfo{pages}{187} (\bibinfo{year}{1976}).

\bibitem[{\citenamefont{Levush et~al.}(1992)\citenamefont{Levush, Antonsen,
  Bromborsky, Lou, and Carmel}}]{LevushBWO}
\bibinfo{author}{\bibfnamefont{B.}~\bibnamefont{Levush}},
  \bibinfo{author}{\bibfnamefont{T.~M.} \bibnamefont{Antonsen}},
  \bibinfo{author}{\bibfnamefont{A.}~\bibnamefont{Bromborsky}},
  \bibinfo{author}{\bibfnamefont{W.~R.} \bibnamefont{Lou}}, \bibnamefont{and}
  \bibinfo{author}{\bibfnamefont{Y.}~\bibnamefont{Carmel}},
  \bibinfo{journal}{IEEE Trans. Plasma Sci.} \textbf{\bibinfo{volume}{20}},
  \bibinfo{pages}{263} (\bibinfo{year}{1992}).

\bibitem[{\citenamefont{Cooke et~al.}(2000)\citenamefont{Cooke, Mondelli,
  Levush, Antonsen, Chernin, McClure, Whaley, and Basten}}]{Space}
\bibinfo{author}{\bibfnamefont{S.~J.} \bibnamefont{Cooke}},
  \bibinfo{author}{\bibfnamefont{A.~A.} \bibnamefont{Mondelli}},
  \bibinfo{author}{\bibfnamefont{B.}~\bibnamefont{Levush}},
  \bibinfo{author}{\bibfnamefont{T.~M.} \bibnamefont{Antonsen}},
  \bibinfo{author}{\bibfnamefont{D.~P.} \bibnamefont{Chernin}},
  \bibinfo{author}{\bibfnamefont{T.~H.} \bibnamefont{McClure}},
  \bibinfo{author}{\bibfnamefont{D.~R.} \bibnamefont{Whaley}},
  \bibnamefont{and} \bibinfo{author}{\bibfnamefont{M.}~\bibnamefont{Basten}},
  \bibinfo{journal}{IEEE Trans. Plasma Sci.} \textbf{\bibinfo{volume}{28}},
  \bibinfo{pages}{841} (\bibinfo{year}{2000}).

\bibitem[{\citenamefont{Brau}(1990)}]{Braubook}
\bibinfo{author}{\bibfnamefont{C.~A.} \bibnamefont{Brau}},
  \emph{\bibinfo{title}{Free-electron laser}} (\bibinfo{publisher}{Academic
  Press, San Diego}, \bibinfo{year}{1990}).

\bibitem[{\citenamefont{Bonifacio et~al.}(1984)\citenamefont{Bonifacio,
  Pellegrini, and Narducci}}]{Collective}
\bibinfo{author}{\bibfnamefont{R.}~\bibnamefont{Bonifacio}},
  \bibinfo{author}{\bibfnamefont{C.}~\bibnamefont{Pellegrini}},
  \bibnamefont{and} \bibinfo{author}{\bibfnamefont{L.~M.}
  \bibnamefont{Narducci}}, \bibinfo{journal}{Opt. Commun.}
  \textbf{\bibinfo{volume}{40}}, \bibinfo{pages}{373} (\bibinfo{year}{1984}).

\end{thebibliography}

\end{document}